\documentclass[a4paper]{article}
\usepackage{array,amsfonts,amsmath,graphicx, xcolor, graphics}
\usepackage{INTERSPEECH2019}
\usepackage{libertine}
\usepackage{multirow,array}
\usepackage[font=small]{caption}
\usepackage{subcaption}

\pdfoutput=1

\usepackage[activate]{microtype}
\title{Augmenting Generative Adversarial Networks for\\ Speech Emotion Recognition}
\name{Siddique Latif$^{1,2}$, Muhammad Asim$^3$,  Rajib Rana$^1$, Sara Khalifa$^{3,4}$,  Raja Jurdak$^5$, Bj\"{o}rn W.\ Schuller$^{6,7}$}
\address{
  $^1$University of Southern Queensland, Australia\\
  $^2$Distributed Sensing Systems Group, Data61, CSIRO Australia\\
  $^3$Information Technology University, Lahore, Pakistan\\
  $^4$University of New South Wales, Australia\\
  $^5$Queensland University of Technology, Australia\\
  $^6$GLAM -- Group on Language, Audio, \& Music, Imperial College London, UK\\
  $^7$Chair of Embedded Intelligence for Health Care and Wellbeing, University of Augsburg, Germany}
\email{siddique.latif@usq.edu.au}

\begin{document}

\maketitle
\begin{abstract}
Generative adversarial networks (GANs) have shown potential in learning emotional attributes and generating new data samples. However, their performance is usually hindered by the unavailability of larger speech emotion recognition (SER) data. In this work,  we propose a framework that utilises the mixup data augmentation scheme to augment the GAN in feature learning and generation. To show the effectiveness of the proposed framework, we present results for SER on (i) synthetic feature vectors, (ii) augmentation of the training data with synthetic features, (iii)  encoded features in compressed representation. Our results show that the proposed framework can effectively learn compressed emotional representations as well as it can generate synthetic samples that help improve performance in within-corpus and cross-corpus evaluation. 
  
\end{abstract}

\noindent\textbf{Index Terms}: speech emotion recognition, mixup, data augmentation, generative 
adversarial networks, feature learning. 

\section{Introduction}
Speech emotion recognition (SER) is an active area of research with potential applications in healthcare \cite{rana2019automated}, call centres \cite{burkhardt2006detecting}, and designing naturalistic voice-based human-computer interfaces \cite{latif2020deep}.  Despite significant progress in machine learning, the performance of state-of-the-art SER systems is quite low. Data scarcity is one of the major reasons in this field \cite{schuller2013paralinguistics}. Available SER datasets are relatively small in size compared to other speech-related applications such as speaker identification and speech recognition \cite{latif2020deep}. This limits the performance of SER systems by causing the curse of the dimensionality problem \cite{bellman2015adaptive}. Dimensionality reduction techniques are considered as a popular solution to resolve this issue \cite{paraskevopoulos2019unsupervised}. However, features extracted in low dimension using these techniques are not always guaranteed to provide the best performance in SER \cite{sahu2017adversarial}.

Another promising approach is to generate synthetic samples using generative models for augmentation of training data. Generative adversarial networks (GANs)~\cite{goodfellow2014generative} have gained a lot of attention in the machine learning (ML) community due to their ability to learn and mimic data distributions. They have shown great performance in image generation \cite{wang2018high}, image translation \cite{yi2017dualgan}, and enhancement \cite{usman2020retrospective}, and also in speech generation \cite{chandna2019wgansing} and conversion \cite{Hsu2017}. However, the lack of availability of larger labelled datasets causes convergence issues in vanilla GANs while generating the synthetic feature vector to augment SER systems \cite{sahu2018enhancing}. To solve this issue, we propose to use a data augmentation technique combined with a GAN to improve the generation of synthetic samples.  Particularly, we utilise a recently proposed data augmentation technique called ``mixup'' \cite{zhang2017mixup} to train a GAN for synthetic emotional feature generation and also for learning compressed emotional representation. To the best
of our knowledge, this paper is the first to investigate mixup to augment GANs.


The key contribution of this paper is the proposed framework that can effectively utilise mixup while training a GAN, which augments the representation learning as well as synthetic feature vector generation by a GAN. We present a detailed analysis by evaluating the SER performance on (i) a compressed representation, (ii) synthetic samples, and (iii) by using generated samples to augment the training data. Results for within-corpus and cross-corpus setting using two emotional datasets show that the proposed framework performs better compared to recent studies.

\section{Related Work}
GANs have already successfully been applied in SER. Bao et al.\ \cite{bao2019cyclegan}, utilised larger unlabelled data in a 
Cycle consistent adversarial networks (CycleGANs) \cite{zhu2017unpaired} based model to generate synthetic features by transferring an emotion feature vector from an unlabelled speech corpus. They were able to improve the SER performance by utilising synthetic data. Sahu et al. \cite{sahu2018enhancing} investigated two networks including vanilla GAN and a conditional GAN to generate a high-dimensional (1582-d) emotional feature vectors from a low-dimensional (2-d) space. They used support vector machines (SVMs) for emotion classification on real and synthetic data. It was shown in \cite{sahu2018enhancing}, that the vanilla GAN could not achieve convergence due to the limited size of data. They were able to generate synthetic feature vectors by conditioning a GAN on class labels. However, the performance on synthetic features vector was quite low. To address this issue, we are using the mixup strategy on the training data to augment generating abilities in the GAN. 

Some studies also utilised generative models for emotional representation learning \cite{Han18-TCA}. Chang and Scherer \cite{chang2017learning} utilised a deep convolutional GAN in a multi-task setting to learn the emotional representation from speech. They utilised unlabelled data in a semi-supervised way to improve the performance of the system. In \cite{latif2019unsupervised},  the authors utilised the GAN based framework for multi-lingual emotion recognition. Based on the results, they showed that a GAN can help in learning language invariant features. To learn emotional features in lower dimensions, the authors in  \cite{sahu2017adversarial} utilised adversarial autoencoders (AAEs) in SER. Based on their results, the authors showed that AAEs can efficiently encode emotional attributes in lower dimensions. Similarly, the authors in \cite{paraskevopoulos2019unsupervised} explored different low-rank representations learning algorithms for SER. They showed that low dimensional emotional representations can achieve comparable performance to the high dimensional features. To further improve performance on compressed features, we utilise a GAN based framework to learn emotional representation from augmented data. Beyond, GANs have in SER also been used on audio-level for augmentation, e.\,g., by emotional voice conversion \cite{Rizos20-SFE}. An overview on GANs in SER is further found in \cite{Han19-ATI}.

The mixup data augmentation strategy has been applied in various vision-related tasks and also in speech-related studies. In \cite{zhu2019mixup}, the authors use mixup strategies in a deep neural network (DNN)-based text-independent speaker verification system. They were able to significantly improve performance while using mixup. Tomashenko et al.\  \cite{tomashenko2018speaker} utilised mixup for regularisation of DNN-based acoustic models in automatic speech recognition (ASR). They found that mixup provides an additional gain in ASR performance. However, no study has utilised mixup in conjunction with GANs to augment feature learning and generation. 

\section{Proposed Framework}
Our proposed framework consists of two components: mixup and GANs. We briefly explain both components first, and then present the details of the proposed technique. 
\subsection{Mixup Augmentation}
Mixup \cite{zhang2017mixup} is a simple data augmentation technique which trains a neural network on convex combinations of pairs of examples and their labels. In this way, it regularises the neural network to favour simple linear behaviour in-between training examples. It constructs virtual training examples as follows:
\begin{equation}
\label{m1}
    \tilde{x}=\lambda x_{i}+(1-\lambda) x_{j}
\end{equation}
\begin{equation}
\label{m2}
    \tilde{y}=\lambda y_{i}+(1-\lambda) y_{j},
\end{equation}
where ($x_{i}$, $y_{i}$) and ($x_{j}$, $y_{j}$) are randomly selected two examples from training data, and $\lambda$ $\in$ [0, 1].
Therefore, mixup extends the training distribution by augmenting the data with linear interpolations of training samples and their targets. Despite the simplicity of mixup, it can improve the performance of various state-of-the-art systems in computer vision and the audio domain \cite{zhu2019mixup}. As outlined, mixup is an essential part of our proposed framework, and it is used to augment the training data.  

\subsection{Generative Adversarial Networks (GANs)}
Generative adversarial networks (GANs) \cite{goodfellow2014generative} include two neural networks---a generator, $G$, and a discriminator, $D$, which play a min-max adversarial game to contest each other. Given a random sample $z$ from some known prior, $p_{z}$ (e.\,g., Gaussian), $G$ is responsible for generating a fake or synthetic data point $G(z)$. The discriminator, $D$, attempts to differentiate between generated samples, $G(z)$, and real data samples, $x$, (drawn from data distribution, $p_{\text{data}}$). The objective of a GAN is to train generator network, $G(z)$ that can mimic real data such that the discriminator becomes incapable of discriminating between real and synthetic samples. This makes the GANs very powerful in feature learning \cite{latif2020deep} and generation \cite{duarte2019wav2pix}. In SER, their performance is hindered by the de facto unavailability of larger datasets. We aim to address this issue by proposing a framework that can utilise mixup in an effective way to augment GANs both in feature learning as well as in feature generation.


\subsection{Augmenting GANs}
\begin{figure}[!t]
  \centering
  \includegraphics[width=\linewidth]{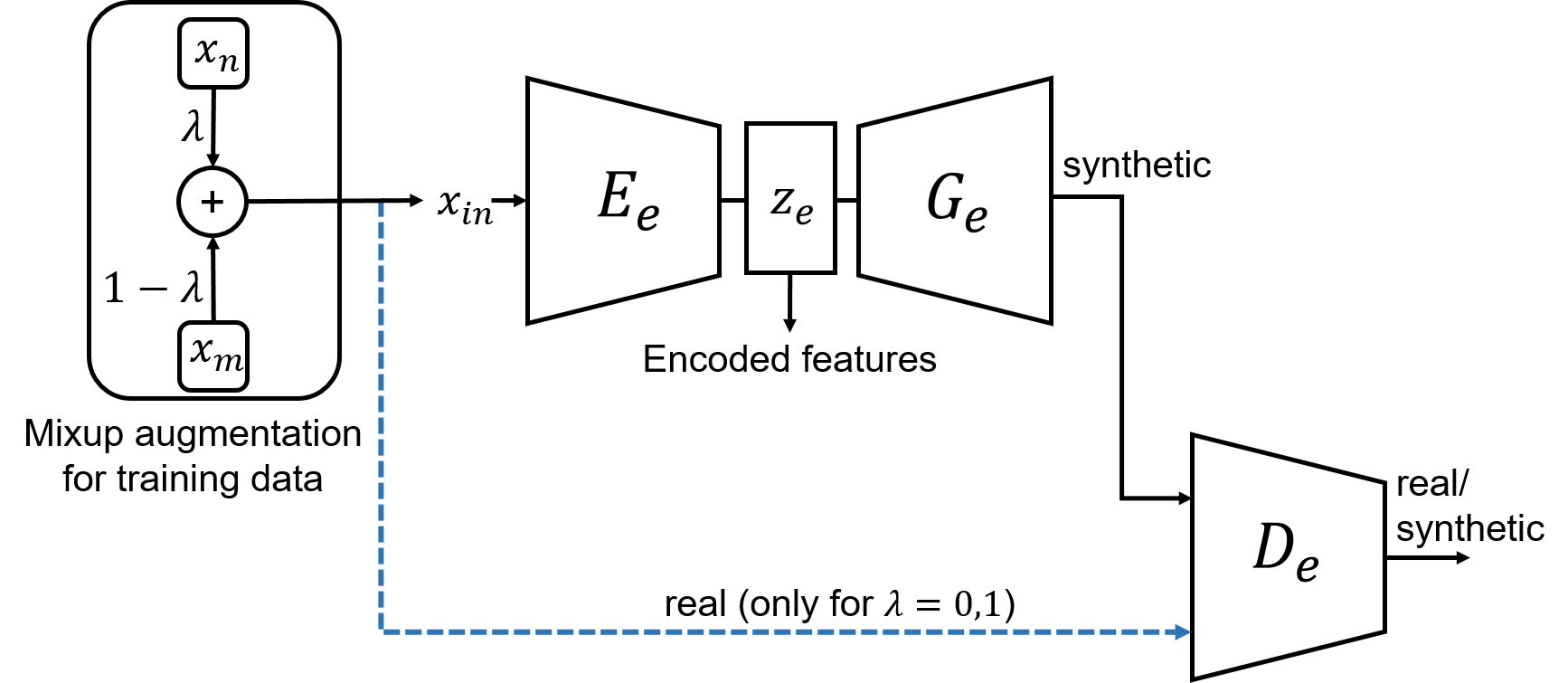}
  \caption{Block diagram of the proposed framework. Mixup is only applied to the training data. The blue doted line shows that $D_{e}$ is only updated for real samples ($\lambda= 0, 1$). }
  \label{fig:model}
\end{figure}

As outlined, 
our model combines mixup with GAN to augment feature learning and generation in SER. The model is shown in Figure \ref{fig:model}. 
We use mixup to linearly interpolate the input samples before providing them to the proposed GAN network. The samples ($x_{n}$, $y_{n}$) and ($x_{m}$, $y_{m}$) are randomly selected from the training data to create mixup samples $x_{in}$ and their labels using the equations \ref{m1} and \ref{m2}. Due to the unsupervised nature of our proposed framework, only $x_{in}$ samples are given to the encoder ($E_{e}$) network. Here, we modified the GAN architecture and use an encoder ($E_{e}$) network along with a generator ($G_{e}$) and a discriminator ($D_{e}$). The encoder network $E_{e}$ generates the compressed encoded feature vector $z_{e}$. Instead of a random sample, the generator ($G_{e}$) uses encoded features $z_{e}$ to generate synthetic (or fake) samples ($G_{e}(z_{e})$). The generator ($G_{e}$) also acts as the decoder of the autoencoder network. The parameters of the encoder and decoder are optimised by minimising the following cost function:
\begin{equation}
\label{AE}
    \mathcal{L}(x_{in},G_{e}(E_{e}(x)))=\lVert{x_{in}-\hat{x}_{in}}\rVert_{2}^{2}.
\end{equation}
The discriminator ($D_{e}$) is tasked to classify between real and synthetic ($G_{e}(z_{e})$) samples using a binary cross-entropy loss function. Here, we consider real samples with $\lambda= 0, 1$ . Therefore, the discriminator ($D_{e}$) network is tasked to classify the real sample with $\lambda= 0, 1$, and the synthetic one. This enables the generator ($G_{e}$) to generate samples close to real samples ($\lambda= 0, 1$) instead of confusions arising from augmented samples with mixup. It also helps the encoder network to encode important emotional attributes that can help $G_{e}$ in synthetic feature generation.  Overall, the proposed model is trained using the following optimisation: 
\begin{equation}
\label{pro}
    \underset{G_{e}}{\text{min}} \  \underset{D_{e}}{\text{max}} \quad \mathrm{E}_x[\log(D_{e}(x))] + \mathrm{E}_y[\log(1 - D_{e}(G_{e}(E_{e}(z_{e}))))].
\end{equation}
The generator ($G_{e}$) attempts to minimise the optimisation in Equation \ref{pro} by generating a synthetic sample that can fool the discriminator in the classification of real samples ($\lambda$= 0, 1) and generated ones.   We train the overall model iteratively. First, we update the autoencoder network.  Then, the generator network is updated. Finally, the discriminator network is updated for samples with $\lambda= 0, 1$.


\begin{figure*}[!ht]%
\centering
\begin{subfigure}{0.25\linewidth}
\includegraphics[trim=0cm 0cm 1.7cm 1.4cm,clip=true,width=\linewidth]{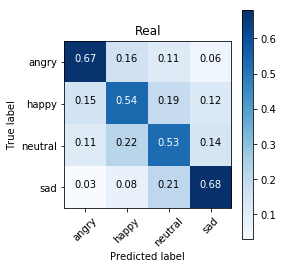}%
\captionsetup{justification=centering}
\caption{real features} %
\label{real}%
\end{subfigure}%
\begin{subfigure}{0.25\linewidth}
\includegraphics[trim=0cm 0cm 1.7cm 1.4cm,clip=true,width=\linewidth]{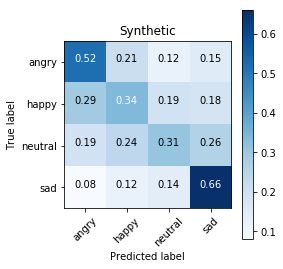}%
\captionsetup{justification=centering}
\caption{synthetic features}%
\label{synd}%
\end{subfigure}
\begin{subfigure}{0.25\linewidth}
\includegraphics[trim=0cm 0cm 1.7cm 1.4cm,clip=true,width=\linewidth]{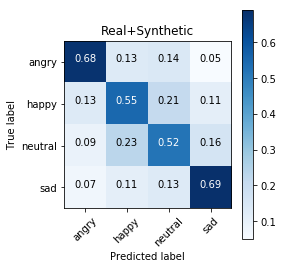}%
\captionsetup{justification=centering}
\caption{real+synthetic} %
\label{real+syn}%
\end{subfigure}%
\begin{subfigure}{0.25\linewidth}
\includegraphics[trim=0cm 0cm 1.7cm 1.4cm,clip=true,width=\linewidth]{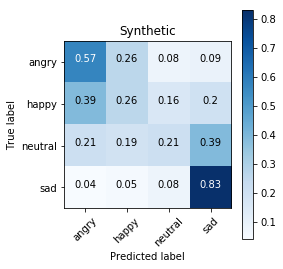}%
\captionsetup{justification=centering}
\caption{synthetic features Bao et al.\ \cite{bao2019cyclegan}} %
\label{bao}%
\end{subfigure}%
\caption{Results on the IEMOCAP data using: (\ref{real}) real features, (\ref{synd}) synthetic features, and (\ref{real+syn}) real+synthetic. \ref{bao} shows the results of \cite{bao2019cyclegan} on synthetic features.}
\label{fig:dataeffect}
\end{figure*}

\section{Experimental Setup}
\subsection{Dataset}
We use the following datasets for evaluations.\linebreak
\textbf{IEMOCAP:} Interactive Emotional Dyadic Motion Capture (IEMOCAP) \cite{busso2008iemocap} is a multimodal corpus that contains English dyadic conversations of ten actors over five sessions. Each session has recordings from one male and one female speaker. Overall, utterances in IEMOCAP are annotated in 10 emotions by 3-4 assessors based on both video and audio streams. To be consistent with previous  studies \cite{Latif2018,latif2018transfer}, we use four emotions including angry, happy, neutral, and sad, where the excitement class is merged into the happiness class. This results in a total of 5\,531 samples.\linebreak
\textbf{MSP-IMPROV:} For cross-corpus evaluation, we select the MSP-IMPROV \cite{busso2017msp} dataset as target data. This corpus also contains the recordings of English dyadic interactions between actors. There are six sessions, where each session has the utterances from two speakers (one male, and one female). Overall, 7\,798 utterances from 12 speakers are annotated across four emotions: angry, neutral, sad, happy. We use all utterances of this corpus.

\subsection{Features}
We use the openSMILE toolkit \cite{eyben2013recent} for extracting features from speech utterances. We use ‘emobase2010’ as reference feature set which consists of 1\,582 features. This feature set is based on the Interspeech 2010 Paralinguistics Challenge feature set (IS10) \cite{schuller2010interspeech} and contains the combination of prosody, spectral, and energy-based features. We use these features as real samples in our experiments. Mixup is applied directly on these features 
of training samples and their labels.

\subsection{Model Configuration and Training}
\label{Model_C}
We implement our model using feed-forward neural network layers. Our encoder and decoder network consist of two layers with hidden units of 1\,000 and 500 each. We vary the dimension of the encoder feature vectors to compare the results with different studies. Our discriminator consists of two hidden layers with 1\,000 neurons each. The autoencoder network is regularised by a dropout layer with a value of 0.5 for between two feed-forward layers. Leaky Rectified Linear Units (leaky ReLUs) \cite{xu2015empirical} are selected as activation function in all hidden layers.  

As described, we employ mixup on the IS10 features vectors and their respective labels of training data. Augmented training data is then given to the proposed model. We pre-train the autoencoder network before initialising the generator. The generator is updated for all input samples, however, the discriminator is only updated for input samples with $\lambda= 0, 1$. It is important to note that we only use mixup on training data. After training the model, we use it to compute the encoded feature vectors and synthetic data for training as well as the testing set. 
We consider utterance-level speaker-independent SER for our experiments. Specifically, we use leave-one-session-out cross-validation to be consistent with previous studies. We use the unweighted average recall (UAR) as the performance metric. We repeat all experiments five times and
mean and standard deviation are reported. We apply min-max normalisation in the synthetic features generation experiments. For cross-corpus evaluation, we apply z-normalisation separately, as it provides better results compared to min-max normalisation for cross-corpus classification \cite{zhang2011unsupervised}.

\section{Experiments and Results}
We perform two types of experiments to evaluate the performance of the proposed framework: (1) a within-corpus experiment, and (2) a cross-corpus experiment.  Each experiment is presented separately below.
\subsection{Within-corpus experiments}
In this experiment, we evaluate the proposed model on both synthetic and encoded features. 
\subsubsection{Synthetic features}
\label{synthfea}

In this experiment, we perform analysis on synthetic features. We build DNN classifiers for emotion classification using: (i) only real features, (ii) only synthetic features, and (iii) both real and synthetic features. Here, real features show the openSMILE ones with the mixup scheme. Our classifiers consist of two hidden layers with 400 hidden units for the experiments (i) and (ii), and 1000 hidden units for the experiment (iii). We use the dropout layer with a dropout value of 0.5. We use a learning rate of $10^{-5}$ in all these experiments. Results are reported in Table \ref{syn}. 

\begin{table}[!ht]
\scriptsize
\caption{Results for cross-validation evaluation on IEMOCAP}
\centering
\begin{tabular}{|l|l|l|l|}
\hline
Studies & Real & Syn. & Real+Syn. \\ \hline
Sahu et al.\ \cite{sahu2018enhancing} & 59.42     & 34.09      & 60.29          \\ \hline
Bao et al.\ \cite{bao2019cyclegan}& 59.48 $\pm$ 0.71     & 46.59 $\pm$ 0.75      &60.37 $\pm$0.70            \\ \hline
Ours                        &\textbf{60.51$\pm$0.57}      &  45.75 $\pm$ 0.81    & \textbf{ 61.05 $\pm$0.68}         \\ \hline
\end{tabular}
\label{syn}
\end{table}

We perform a comparison of our results to recent studies \cite{bao2019cyclegan} and \cite{sahu2018enhancing}.  In \cite{sahu2018enhancing}, Sahu et al.\ investigated GAN architectures to generate the synthetic feature vectors (1582-d) using a low dimensional (2-d) representation for SER and to improve the performance in exploiting both real and synthetic features. Similar to \cite{sahu2018enhancing}, we also select $z_{e}=2$ and generate a synthetic vector (1582-d). We are achieving better results compared to this study for the classification of real, synthetic, and real+synthetic settings. Bao et al.\ \cite{bao2019cyclegan} apply a CycleGAN based model to augment SER by transferring feature vectors extracted from a large unlabelled speech data into synthetic features for the given target emotions. We compare their best results for real+synthetic features when they used the classification loss in Table \ref{syn}. In contrast to \cite{bao2019cyclegan}, we are achieving better results for real and real+synthetic features. However, our classification results on synthetic features are slightly lower. To gain a deeper understanding of the performance differences, we analyse prediction errors in Figure \ref{real}-\ref{real+syn}. We also plot the prediction results on synthetic data achieved by \cite{bao2019cyclegan} in Figure \ref{bao}. 

It can be noted from the confusion matrices that the prediction performance is improved using real+synthetic features compared to using only real features (see Figure \ref{real} and \ref{real+syn}). Our results on synthetic data (Figure \ref{synd}) are comparable to the results achieved using real data (Figure \ref{real}) for the angry and sad classes. However, we are achieving lower results for the classes happy and neutral.  We also compare the prediction errors on synthetic data with Bao et al.\ \cite{bao2019cyclegan}. 
The proposed model in \cite{bao2019cyclegan} improved the prediction on the sad class, however, performed poor on happy and neutral (see Figure \ref{bao}). In contrast, we are achieving results closer to Figure  \ref{real} for all classes, which shows that the proposed framework is generating synthetic feature vectors similar to real samples.

\subsubsection{Encoded Features}

To evaluate the performance of encoded features by our proposed model, we use encoded features ($z_{e}=E_{e}(x_{in})$) from the autoencoder component as the input to the classifier for classification. In this experiment, we compare our results with a recent study \cite{paraskevopoulos2019unsupervised} in which the authors used different non-linear dimensionality reduction algorithms for extracting low-rank feature representations for SER. We select three top-performing dimensionality reduction algorithms in \cite{paraskevopoulos2019unsupervised}. These methods include SMACOF multidimensional scaling (MDS) \cite{torgerson1952multidimensional}, Principal Component Analysis (PCA) \cite{pearson1901liii}, and an autoencoder \cite{hinton1994autoencoders}. Results are presented in Table \ref{table:nls}.

\begin{table}[!ht]
\centering
\scriptsize
\caption{Comparison of results using different dimensionality reduction algorithms on IEMOCAP. }
\begin{tabular}{|l|l|}
\hline
Method         & UAR (\%)        \\ \hline
SMACOF MDS \cite{paraskevopoulos2019unsupervised} &  58.5               \\ \hline
 PCA \cite{paraskevopoulos2019unsupervised}&  57.7              \\ \hline
Autoencoder \cite{paraskevopoulos2019unsupervised} & 57.8                \\ \hline
\multicolumn{2}{|c|}{with mixup} \\ \hline
 SMACOF MDS  & 58.9               \\ \hline
 PCA &         58.3        \\ \hline
Autoencoder&   58.5              \\ \hline
Proposed &    \textbf{59.6}             \\ \hline
\end{tabular}
\label{table:nls}
\end{table}

In \cite{paraskevopoulos2019unsupervised}, the authors used SVMs for classification on the features learnt by each dimensionality reduction algorithm. However, they did not use any data augmentation technique. Therefore, we also implemented these dimensionality reduction methods with mixup to have a fair comparison with our proposed model. To be consistent with \cite{paraskevopoulos2019unsupervised}, we reduce the dimension of the IS10 features from 1\,582 to 25 dimensions and compute the results. In our proposed model, we use $z_{e}=E_{e}(x_{in})$ features for classification with SVMs. We select an RBF kernel and perform a grid search on validation data to select the optimal hyper-parameters for classification. The standard autoencoder applied in this experiment is trained with 3 fully connected encoder layers, 3 decoder layers and 1 hidden layer. ReLU activation is chosen in these layers. It can be noted from Table \ref{table:nls} that the proposed model performs better than the other non-linear dimension reduction techniques. This shows that the proposed model efficiently encodes features in lower dimension while keeping emotional information.

\subsection{Cross-corpus evaluation}
To investigate the proposed model in a cross-corpus setting, we also perform the same experiments (as in Section \ref{synthfea}) using real, synthetic, and real+ synthetic data. Here, we have MSP-IMPROV as the target data. Therefore, we randomly select 30\,\% of the samples from MSP-IMPROV as the development set for hyper-parameter selection and the remaining 70\,\% as test data, as done in \cite{bao2019cyclegan}. We keep the class proportions equal in both sets. For classification, we choose a DNN model with two fully connected layers with 400 hidden units in each layer. The values for the learning rate and dropout are $10^{-5}$ and $0.8$, respectively.

\begin{table}[!ht]
\scriptsize
\caption{Results for cross-corpus evaluation.}
\centering
\begin{tabular}{|l|l|l|l|}
\hline
Studies & Real & Syn. & Real+Syn. \\ \hline
Sahu et al.\ \cite{sahu2018enhancing} & 45.14     & 33.96      & 45.40          \\ \hline
Bao et al.\ \cite{bao2019cyclegan}& 45.58 $\pm$ 0.40     & 41.58 $\pm$ 1.29 & 46.52$\pm$0.43          \\ \hline
Ours                        &\textbf{46.0$\pm$0.57}      &  \textbf{42.15 $\pm$ 1.12}    &  \textbf{46.60 $\pm$0.45}         \\ \hline
\end{tabular}
\label{cross}
\end{table}
The results are compared with \cite{sahu2018enhancing} and \cite{bao2019cyclegan} in Table \ref{cross}. Both of these studies augmented the training data with synthetic samples to help SER in a cross-corpus setting. Compared to these studies, we are achieving improved results. This shows that the proposed model improves the performance of SER in a cross-corpus setting using synthetic data and also when training data is augmented with these synthetic samples.


\section{Conclusions}
A major challenge in speech emotion recognition (SER) is the lack of availability of larger datasets. This limits the performance of representation learning algorithms and generative models. 
We address this issue by proposing a framework that utilises a data augmentation technique called mixup to augment GANs in representation learning as well as synthetic feature vector generation. Compared to recent studies, our proposed framework was able to learn better emotional representations in compressed form and also to generate synthetic features vectors that can be effectively utilised to augment the training size of SER for performance improvement. In future efforts, we aim to design an extended version of the proposed framework for domain adaptation in cross-lingual SER.





\begin{thebibliography}{10}
\providecommand{\url}[1]{#1}
\csname url@samestyle\endcsname
\providecommand{\newblock}{\relax}
\providecommand{\bibinfo}[2]{#2}
\providecommand{\BIBentrySTDinterwordspacing}{\spaceskip=0pt\relax}
\providecommand{\BIBentryALTinterwordstretchfactor}{4}
\providecommand{\BIBentryALTinterwordspacing}{\spaceskip=\fontdimen2\font plus
\BIBentryALTinterwordstretchfactor\fontdimen3\font minus
  \fontdimen4\font\relax}
\providecommand{\BIBforeignlanguage}[2]{{%
\expandafter\ifx\csname l@#1\endcsname\relax
\typeout{** WARNING: IEEEtran.bst: No hyphenation pattern has been}%
\typeout{** loaded for the language `#1'. Using the pattern for}%
\typeout{** the default language instead.}%
\else
\language=\csname l@#1\endcsname
\fi
#2}}
\providecommand{\BIBdecl}{\relax}
\BIBdecl

\bibitem{rana2019automated}
R.~Rana, S.~Latif, R.~Gururajan, A.~Gray, G.~Mackenzie, G.~Humphris, and
  J.~Dunn, ``Automated screening for distress: A perspective for the future,''
  \emph{European Journal of Cancer Care}, p. e13033, 2019.

\bibitem{burkhardt2006detecting}
F.~Burkhardt, J.~Ajmera, R.~Englert, J.~Stegmann, and W.~Burleson, ``Detecting
  anger in automated voice portal dialogs,'' in \emph{Ninth International
  Conference on Spoken Language Processing}, 2006.

\bibitem{latif2020deep}
S.~Latif, R.~Rana, S.~Khalifa, R.~Jurdak, J.~Qadir, and B.~W. Schuller, ``Deep
  representation learning in speech processing: Challenges, recent advances,
  and future trends,'' \emph{arXiv preprint arXiv:2001.00378}, 2020.

\bibitem{schuller2013paralinguistics}
B.~Schuller, S.~Steidl, A.~Batliner, F.~Burkhardt, L.~Devillers, C.~M{\"u}Ller,
  and S.~Narayanan, ``Paralinguistics in speech and language—state-of-the-art
  and the challenge,'' \emph{Computer Speech \& Language}, vol.~27, no.~1, pp.
  4--39, 2013.

\bibitem{bellman2015adaptive}
R.~E. Bellman, \emph{Adaptive control processes: a guided tour}.\hskip 1em plus
  0.5em minus 0.4em\relax Princeton university press, 2015.

\bibitem{paraskevopoulos2019unsupervised}
G.~Paraskevopoulos, E.~Tzinis, N.~Ellinas, T.~Giannakopoulos, and
  A.~Potamianos, ``Unsupervised low-rank representations for speech emotion
  recognition,'' \emph{Proc. Interspeech 2019}, pp. 939--943, 2019.

\bibitem{sahu2017adversarial}
S.~Sahu, R.~Gupta, G.~Sivaraman, W.~AbdAlmageed, and C.~Espy-Wilson,
  ``Adversarial auto-encoders for speech based emotion recognition,''
  \emph{Proc. Interspeech 2017}, pp. 1243--1247, 2017.

\bibitem{goodfellow2014generative}
I.~Goodfellow, J.~Pouget-Abadie, M.~Mirza, B.~Xu, D.~Warde-Farley, S.~Ozair,
  A.~Courville, and Y.~Bengio, ``Generative adversarial nets,'' in
  \emph{Advances in neural information processing systems}, 2014, pp.
  2672--2680.

\bibitem{wang2018high}
T.-C. Wang, M.-Y. Liu, J.-Y. Zhu, A.~Tao, J.~Kautz, and B.~Catanzaro,
  ``High-resolution image synthesis and semantic manipulation with conditional
  gans,'' in \emph{Proceedings of the IEEE conference on computer vision and
  pattern recognition}, 2018, pp. 8798--8807.

\bibitem{yi2017dualgan}
Z.~Yi, H.~Zhang, P.~Tan, and M.~Gong, ``Dualgan: Unsupervised dual learning for
  image-to-image translation,'' in \emph{Proceedings of the IEEE international
  conference on computer vision}, 2017, pp. 2849--2857.

\bibitem{usman2020retrospective}
M.~Usman, S.~Latif, M.~Asim, B.-D. Lee, and J.~Qadir, ``Retrospective motion
  correction in multishot mri using generative adversarial network,''
  \emph{Scientific Reports}, vol.~10, no.~1, pp. 1--11, 2020.

\bibitem{chandna2019wgansing}
P.~Chandna, M.~Blaauw, J.~Bonada, and E.~G{\'o}mez, ``Wgansing: A multi-voice
  singing voice synthesizer based on the wasserstein-gan,'' in \emph{2019 27th
  European Signal Processing Conference (EUSIPCO)}.\hskip 1em plus 0.5em minus
  0.4em\relax IEEE, 2019, pp. 1--5.

\bibitem{Hsu2017}
C.-C. Hsu, H.-T. Hwang, Y.-C. Wu, Y.~Tsao, and H.-M. Wang, ``Voice conversion
  from unaligned corpora using variational autoencoding wasserstein generative
  adversarial networks,'' in \emph{Proc. Interspeech 2017}, 2017, pp.
  3364--3368.

\bibitem{sahu2018enhancing}
S.~Sahu, R.~Gupta, and C.~Espy-Wilson, ``On enhancing speech emotion
  recognition using generative adversarial networks,'' \emph{Proc. Interspeech
  2018}, pp. 3693--3697, 2018.

\bibitem{zhang2017mixup}
H.~Zhang, M.~Cisse, Y.~N. Dauphin, and D.~Lopez-Paz, ``mixup: Beyond empirical
  risk minimization,'' \emph{arXiv preprint arXiv:1710.09412}, 2017.

\bibitem{bao2019cyclegan}
F.~Bao, M.~Neumann, and N.~T. Vu, ``Cyclegan-based emotion style transfer as
  data augmentation for speech emotion recognition,'' \emph{Manuscript
  submitted for publication}, pp. 35--37, 2019.

\bibitem{zhu2017unpaired}
J.-Y. Zhu, T.~Park, P.~Isola, and A.~A. Efros, ``Unpaired image-to-image
  translation using cycle-consistent adversarial networks,'' in
  \emph{Proceedings of the IEEE international conference on computer vision},
  2017, pp. 2223--2232.

\bibitem{Han18-TCA}
J.~Han, Z.~Zhang, Z.~Ren, F.~Ringeval, and B.~Schuller, ``{Towards Conditional
  Adversarial Training for Prediciting Emotions from Speech},'' in
  \emph{{Proceedings ICASSP}}.\hskip 1em plus 0.5em minus 0.4em\relax Calgary,
  Canada: IEEE, April 2018, pp. 6822--6826.

\bibitem{chang2017learning}
J.~Chang and S.~Scherer, ``Learning representations of emotional speech with
  deep convolutional generative adversarial networks,'' in \emph{2017 IEEE
  International Conference on Acoustics, Speech and Signal Processing
  (ICASSP)}.\hskip 1em plus 0.5em minus 0.4em\relax IEEE, 2017, pp. 2746--2750.

\bibitem{latif2019unsupervised}
S.~Latif, J.~Qadir, and M.~Bilal, ``Unsupervised adversarial domain adaptation
  for cross-lingual speech emotion recognition,'' in \emph{2019 8th
  International Conference on Affective Computing and Intelligent Interaction
  (ACII)}.\hskip 1em plus 0.5em minus 0.4em\relax IEEE, 2019, pp. 732--737.

\bibitem{Rizos20-SFE}
G.~Rizos, A.~Baird, M.~Elliott, and B.~Schuller, ``{StarGAN for Emotional
  Speech Conversion: Validated by Data Augmentation of End-to-End Emotion
  Recognition},'' in \emph{{Proceedings ICASSP}}.\hskip 1em plus 0.5em minus
  0.4em\relax Barcelona, Spain: IEEE, 2020.

\bibitem{Han19-ATI}
J.~Han, Z.~Zhang, N.~Cummins, and B.~Schuller, ``{Adversarial Training in
  Affective Computing and Sentiment Analysis: Recent Advances and
  Perspectives},'' \emph{IEEE Computational Intelligence Magazine, Special
  Issue on Computational Intelligence for Affective Computing and Sentiment
  Analysis}, vol.~14, no.~2, pp. 68--81, 2019.

\bibitem{zhu2019mixup}
Y.~Zhu, T.~Ko, and B.~Mak, ``Mixup learning strategies for text-independent
  speaker verification,'' \emph{Proc. Interspeech 2019}, pp. 4345--4349, 2019.

\bibitem{tomashenko2018speaker}
N.~A. Tomashenko, Y.~Y. Khokhlov, and Y.~Est{\`e}ve, ``Speaker adaptive
  training and mixup regularization for neural network acoustic models in
  automatic speech recognition.'' in \emph{Interspeech}, 2018, pp. 2414--2418.

\bibitem{duarte2019wav2pix}
A.~Duarte, F.~Roldan, M.~Tubau, J.~Escur, S.~Pascual, A.~Salvador, E.~Mohedano,
  K.~McGuinness, J.~Torres, and X.~Giro-i Nieto, ``Wav2pix: speech-conditioned
  face generation using generative adversarial networks,'' in \emph{IEEE
  International Conference on Acoustics, Speech and Signal Processing
  (ICASSP)}, vol.~3, 2019.

\bibitem{busso2008iemocap}
C.~Busso, M.~Bulut, C.-C. Lee, A.~Kazemzadeh, E.~Mower, S.~Kim, J.~N. Chang,
  S.~Lee, and S.~S. Narayanan, ``Iemocap: Interactive emotional dyadic motion
  capture database,'' \emph{Language resources and evaluation}, vol.~42, no.~4,
  p. 335, 2008.

\bibitem{Latif2018}
S.~Latif, R.~Rana, J.~Qadir, and J.~Epps, ``Variational autoencoders for
  learning latent representations of speech emotion: A preliminary study,'' in
  \emph{Proc. Interspeech 2018}, 2018, pp. 3107--3111.

\bibitem{latif2018transfer}
S.~Latif, R.~Rana, S.~Younis, J.~Qadir, and J.~Epps, ``Transfer learning for
  improving speech emotion classification accuracy,'' \emph{Proc. Interspeech
  2018}, pp. 257--261, 2018.

\bibitem{busso2017msp}
C.~Busso, S.~Parthasarathy, A.~Burmania, M.~AbdelWahab, N.~Sadoughi, and E.~M.
  Provost, ``Msp-improv: An acted corpus of dyadic interactions to study
  emotion perception,'' \emph{IEEE Transactions on Affective Computing},
  vol.~8, no.~1, pp. 67--80, 2017.

\bibitem{eyben2013recent}
F.~Eyben, F.~Weninger, F.~Gross, and B.~Schuller, ``Recent developments in
  opensmile, the munich open-source multimedia feature extractor,'' in
  \emph{Proceedings of the 21st ACM international conference on
  Multimedia}.\hskip 1em plus 0.5em minus 0.4em\relax ACM, 2013, pp. 835--838.

\bibitem{schuller2010interspeech}
B.~Schuller, S.~Steidl, A.~Batliner, F.~Burkhardt, L.~Devillers, C.~M{\"u}ller,
  and S.~S. Narayanan, ``The interspeech 2010 paralinguistic challenge,'' in
  \emph{Eleventh Annual Conference of the International Speech Communication
  Association}, 2010.

\bibitem{xu2015empirical}
B.~Xu, N.~Wang, T.~Chen, and M.~Li, ``Empirical evaluation of rectified
  activations in convolutional network,'' \emph{arXiv preprint
  arXiv:1505.00853}, 2015.

\bibitem{zhang2011unsupervised}
Z.~Zhang, F.~Weninger, M.~W{\"o}llmer, and B.~Schuller, ``Unsupervised learning
  in cross-corpus acoustic emotion recognition,'' in \emph{2011 IEEE Workshop
  on Automatic Speech Recognition \& Understanding}.\hskip 1em plus 0.5em minus
  0.4em\relax IEEE, 2011, pp. 523--528.

\bibitem{torgerson1952multidimensional}
W.~S. Torgerson, ``Multidimensional scaling: I. theory and method,''
  \emph{Psychometrika}, vol.~17, no.~4, pp. 401--419, 1952.

\bibitem{pearson1901liii}
K.~Pearson, ``{LIII}. on lines and planes of closest fit to systems of points
  in space,'' \emph{The London, Edinburgh, and Dublin Philosophical Magazine
  and Journal of Science}, vol.~2, no.~11, pp. 559--572, 1901.

\bibitem{hinton1994autoencoders}
G.~E. Hinton and R.~S. Zemel, ``Autoencoders, minimum description length and
  helmholtz free energy,'' in \emph{Advances in neural information processing
  systems}, 1994, pp. 3--10.

\end{thebibliography}



\end{document}